\documentclass[11pt]{article}

\usepackage{amsfonts}
\usepackage{amssymb}
\usepackage{amscd}
\usepackage{a4wide}
\usepackage{amsmath,amssymb,amsfonts,amsthm}
\usepackage[all]{xy}
\usepackage{color}

\def\beq{\begin{equation}}
\def\eeq{\end{equation}}
\def\bea{\begin{eqnarray}}
\def\eea{\end{eqnarray}}
\def\nn{\nonumber}

\date{}

\begin{document}

\title{\Large\bf $p$-Adic Structure of the Genetic Code }

\author{Branko
Dragovich\,\footnote{\textsf{\, E-mail:\,dragovich@ipb.ac.rs}}\\
{} \\ {Institute of Physics}
\\ { Pregrevica 118, 11080  Belgrade, Serbia}}

\maketitle


\begin{abstract}
The genetic code is connection between $64$ codons, which are
building blocks of the genes, and $20$ amino acids, which are
building blocks of the proteins. In addition to coding amino
acids, a few codons code stop signal, which is at the end of
genes, i.e. it terminates process of protein synthesis. This
article is a review of simple modelling of the genetic code and
related subjects by concept of $p$-adic distance. It also contains
some new results. In particular, the article presents appropriate
structure of the codon space, degeneration and possible evolution
of the genetic code. $p$-Adic modelling of the genetic code is
viewed as the first step in further application of $p$-adic tools
in the information sector of life science.
\end{abstract}


\section{Introduction}

Francis Crick (1916--2004), who together with James Watson
discovered double helicoidal structure of DNA, in 1953 announced
``We have discovered the secret of life'' (Hayes, 1998). However,
the life has still many secrets and  the genetic code seems to be
the most intriguing one. Although the standard genetic code was
finally experimentally deciphered in 1966, its theoretical
understanding has remained unsatisfactory and new models have been
proposed occasionally. The genetic code is still subject of some
investigations from mathematical, physical, chemical, biological
and bioinformation point of view. However, many of these models
are rather complicated and do not give complete description and
understanding of the various properties of the genetic code.

It is instructive to recall discovery of quantum mechanics. Before
its emergence, many physical experimental data could not be well
described by classical methods. It was necessary to invent new
appropriate physical concepts and to use suitable new mathematical
methods. It seems that a similar situation should happen in
theoretical description of living processes in biological
organisms. To this end, $p$-adic methods seem to be very promising
tools in further investigation of the life.

In this article we  emphasize the role of $p$-adic distance.
Namely, some parts of a biological system can be considered
simultaneously with respect to different metrics -- the usual
Euclidean metric, which measures spatial distances, and some other
metrics, which measure nearness related to some bioinformation (or
other) properties. Here we consider  the genetic code using an
ultrametric space, which elements are codons presented with some
natural numbers and the distance between them is the $p$-adic one.
An ultrametric space $\mathbb{M}$ is a metric space which
distances satisfy strong triangle inequality (also called
ultrametric inequality), i.e.
\[ d(x,y)\, \leq \, \text{max}\{ d(x,z),\, d(z,y) \}\]
for any $x, y, z \in \mathbb{M}.$ The ultrametric inequality was
formulated by Felix Hausdorff in 1934   and ultrametric spaces
were introduced by Marc Krasner in 1944. Ultrametrics is also
named non-Archimedean metrics. Ultrametric spaces  exhibit some
exotic properties. The first application of ultrametricity  was in
biological taxonomy. Ultrametricity in pphysics (Rammal {\it et
al.}, 1986) was observed in 1984 in the context of the mean field
theory of spin glasses and it induced a considerable research  in
many scientific fields (e.g. statistical physics, neural networks,
conformational structure of proteins, diffusion processes,
hierarchical systems). 

Modelling   the genetic code   is   an opportunity for application
of $p$-adic distance. In 2006
 we introduced (Dragovich B. and Dragovich A., 2006) a $p$-adic approach to DNA and RNA
sequences, and to the genetic code. The central point of our
approach is an appropriate identification of four nucleotides with
digits $1,\, 2, \, 3, \, 4$ of $5$-adic number expansions and
application of $p$-adic distances between obtained numbers.
$5$-Adic numbers with three digits form $64$ integers which
correspond to  $64$ codons. In (Dragovich B. and Dragovich A.,
2007) we analyzed $p$-adic degeneracy of the genetic code. As one
of the main results that we have obtained is explanation of the
structure of the genetic code degeneracy using $p$-adic distance
between codons. Paper (Dragovich B. and Dragovich A., 2010)
contains consideration of possible evolution of the genetic code
and some generalizations of $p$-adic modelling of the genetic
code. Article (Dragovich, 2009) is related to the role of number
theory in modelling the genetic code. A similar approach to the
genetic code was reconsidered on diadic plane (Khrennikov and
Kozyrev, 2007).

$p$-Adic models in mathematical physics have been actively
considered since 1987 (see (Brekke {\it et al.}, 1993; Vladimirov
{\it et al.}, 1994) for early reviews and (Dragovich, 2004;
Dragovich, 2006; Dragovich {\it et al.}, 2009) for some recent
reviews). It is worth noting that $p$-adic models with
pseudodifferential operators have been successfully applied to
interbasin kinetics of proteins (Avetisov {\it et al.}, 2002).
Some $p$-adic aspects of cognitive, psychological and social
phenomena have been also considered (Khrennikov, 2004).

To have a self-contained and comprehensible exposition of the
genetic code, we shall first briefly review some basic notions
from molecular biology.

\section{Basic Notions of the Genomics and Proteomics}

One of the essential characteristics that differ a living organism
from all other material systems is related to its genome. The
genome of an organism is its whole hereditary information encoded
in  the desoxyribonucleic acid (DNA), and contains  both coding
and non-coding sequences. In some viruses, which are between
living and non-living objects, genetic material is encoded in the
ribonucleic acid (RNA). Investigation of the entire genome is the
subject of genomics. The human genome is composed of more than
three billion DNA base pairs and its $97$\% is non-coding.

The DNA  is a macromolecule composed of two polynucleotide chains
with a double-helical structure. Nucleotides consist of a base, a
sugar and a phosphate group. The sugar and phosphate groups
provide helical backbone. There are four bases and they are
building elements of the genetic information. They are named
adenine (A), guanine (G), cytosine (C) and thymine (T). Adenine
and guanine are purines, while cytosine and thymine are
pyrimidines. In the sense of information, the nucleotide and its
base present the same object. Nucleotides are arranged along
chains of double helix through base pairs A-T and C-G bonded by 2
and 3 hydrogen bonds, respectively. As a consequence of this
pairing there is an equal number of cytosine and guanine as well
as the equal rate of adenine and thymine. DNA is packaged in
chromosomes which are localized in the nucleus of the eukaryotic
cells.

\begin{table}
{{\bf Table 1.}  List of 20 standard amino acids used in proteins
by living cells.
 3-Letter and 1-letter abbreviations of amino acids,
their chemical structure of side chains, polarity and
hydrophobicity are presented.} \vskip3mm
 \label{Tab:01}
\centerline{ {\begin{tabular}{|l|l|l|l|l|}
 \hline \ & \ & \  & \  &  \\
Amino acids & Abbreviations & Side Chain (R) & Polar & Hydrophobic \\
 \hline \ & \  & \  & \  & \\
 Alanine &  Ala, A  & -$CH_3$ & no & yes \\
 Cysteine &  Cys, C  & -$CH_2$SH & no & yes \\
 Aspartate &  Asp, D  & -$CH_2$COOH  & yes & no \\
 Glutamate &  Glu, E  & -$(CH_2)_2COOH$ & yes & no \\
 Phenynalanine&  Phe, F  & -$CH_2C_6H_5$  & no & yes  \\
 Glycine &  Gly, G  & -$H$ & no & yes \\
 Histidine &  His, H  & -$CH_2$-$C_3H_3N_2$ & yes & no  \\
 Isoleucine &  Ile, I  & -$CH(CH_3)CH_2CH_3$ & no & yes   \\
 Lysine &  Lys, K  & -$(CH_2)_4NH_2$ & yes & no \\
 Leucine &  Leu, L  & -$CH_2CH(CH_3)_2$ & no & yes \\
 Methionine & Met, M  & -$(CH_2)_2SCH_3$ & no & yes\\
 Asparagine & Asn, N  & -$CH_2CONH_2$ & yes & no \\
 Proline &  Pro, P  & -$(CH_2)_3$- & yes & no \\
 Glutamine &  Gln, Q  & -$(CH_2)_2CONH_2$  & yes & no \\
 Arginine &  Arg, R  & -$(CH_2)_3NHC(NH)NH_2$ & yes & no \\
 Serine &  Ser, S  & -$CH_2OH$  & yes & no\\
 Threonine & Thr, T  & -$CH(OH)CH_3$ & yes & no \\
 Valine &  Val, V  & -$CH(CH_3)_2$ & no & yes \\
 Tryptophan &  Trp, W  & -$CH_2C_8H_6N$ & no & yes \\
 Tyrosine &  Tyr, Y   & -$CH_2$-$C_6H_4OH$ & yes & yes \\
 \hline
\end{tabular}}{}}
\vskip3mm
\end{table}

The main role of  DNA is to storage genetic information and there
are two main processes to exploit this information. The first one
is replication, in which  DNA duplicates giving two new DNA
containing the same information as the original one. This is
possible owing to the fact that each of two chains contains
complementary bases of the other one. The second process is
related to the gene expression, i.e. the passage of DNA gene
information to proteins. It performs by the messenger ribonucleic
acid (mRNA), which is usually a single polynucleotide chain. The
mRNA is synthesized during the first part of this process, known
as transcription, when nucleotides C, A, T, G from DNA are
respectively transcribed into their complements G, U, A, C in
mRNA, where T is replaced by U (U is the uracil, which is a
pyrimidine). The next step in gene expression is translation, when
the information coded by codons in the mRNA  is translated into
proteins. In this process  participate also  transfer tRNA and
ribosomal rRNA.

Protein synthesis in all eukaryotic cells performs in the
ribosomes of the cytoplasm. Proteins (Finkelstein and Ptitsyn,
2002) are organic macromolecules composed of amino acids arranged
in a linear chain. The sequence of amino acids in a protein is
determined by sequence of codons contained in RNA genes. Amino
acids are molecules that consist of amino, carboxyl and R (side
chain) groups. Depending on R group there are 20 standard amino
acids. These amino acids are joined together by a peptide bond.
Proteins are substantial ingredients of all living organisms
participating in various processes in cells and determining the
phenotype of an organism. There are more proteins than genes in
DNA, because of alternative splicing of genes and translational
modifications.  In the human body there may be about 2 million
different proteins. The study of proteins, especially their
structure and functions, is called proteomics. The complete
proteome is the entire set of proteins in an organism.

Some properties of amino acids are presented in Table 1. For a
more detailed and comprehensive information on genomics and
proteomics one can use book (Watson {\it et al.}, 2004) on
molecular biology.

\section{General Features of the Genetic Code}

Experimental study of the connection between ordering of
nucleotides in DNA (and RNA) and ordering of amino acids in
proteins led to the  deciphering of the standard genetic code in
the mid-1960s. The genetic code is understood as a dictionary for
translation of codons from  DNA (and RNA) to amino acids during
synthesis of proteins. The information on amino acids is contained
in codons: each codon codes either an amino acid or termination
signal (see, e.g. Table 2 as a standard table of the vertebrate
mitochondrial genetic code). To the sequence of codons in  RNA
corresponds quite definite sequence of amino acids in a protein,
and this sequence of amino acids determines primary structure of
the protein.  At the time of deciphering, it was mainly believed
that the standard code is unique, result of a chance and fixed a
long time ego. Crick (Crick, 1968) expressed such belief in his
"frozen accident" hypothesis, which has not been supported by
later observations. Moreover, it has been discovered so far about
 20 different genetic codes.  However, differences are not drastic and
many common general properties have been found: four nucleotides,
trinucleotide codons, the same mechanism of proton synthesis, ...
 At the first glance the genetic code looks rather arbitrary, but it
is not. Namely, mutations between synonymous codons give the same
amino acid. When mutation alter an amino acid then it is like
substitution of the original by similar one. In this respect the
code is almost optimal.

\begin{table}
{{\bf Table 2.}   The standard (Watson-Crick) table of the
vertebrate mitochondrial genetic code. Ter denotes the terminal
(stop) signal. \vskip3mm \label{Tab:04}}
\centerline{{\begin{tabular}{|l|l|l|l|}
 \hline \ & \ & \ & \\
  UUU \, Phe &   UCU \, Ser &  UAU \, Tyr &  UGU \, Cys  \\
  UUC \, Phe &   UCC \, Ser &  UAC \, Tyr &  UGC \, Cys  \\
  UUA \, Leu &   UCA \, Ser &  UAA \, Ter &  UGA \, Trp  \\
  UUG \, Leu &   UCG \, Ser &  UAG \, Ter &  UGG \, Trp  \\
 \hline \  & \  &  \ & \ \\
  CUU \, Leu &   CCU \, Pro &  CAU \, His &  CGU \, Arg   \\
  CUC \, Leu &   CCC \, Pro &  CAC \, His &  CGC \, Arg   \\
  CUA \, Leu &   CCA \, Pro &  CAA \, Gln &  CGA \, Arg   \\
  CUG \, Leu &   CCG \, Pro &  CAG \, Gln &  CGG \, Arg   \\
 \hline \  & \  & \  &   \\
  AUU \, Ile &   ACU \, Thr &  AAU \, Asn &  AGU \, Ser  \\
  AUC \, Ile &   ACC \, Thr &  AAC \, Asn &  AGC \, Ser  \\
  AUA \, Met &   ACA \, Thr &  AAA \, Lys &  AGA \, Ter  \\
  AUG \, Met &   ACG \, Thr &  AAG \, Lys &  AGG \, Ter  \\
 \hline \ & \   & \  &   \\
  GUU \, Val &   GCU \, Ala  &  GAU \, Asp &  GGU \, Gly  \\
  GUC \, Val &   GCC \, Ala  &  GAC \, Asp &  GGC \, Gly  \\
  GUA \, Val &   GCA \, Ala  &  GAA \, Glu &  GGA \, Gly  \\
  GUG \, Val &   GCG \, Ala  &  GAG \, Glu &  GGG \, Gly  \\
\hline
\end{tabular}}{}}
\end{table}

The relation between codons, on the one hand, and amino acids and
stop signal, from the other hand, is known as the {\it genetic
code}.

Codons are ordered triples composed of C, A, U (T) and G
nucleotides. Each codon presents an information which controls use
of one of the 20 standard amino acids or stop signal in synthesis
of proteins. It is obvious that there are $4 \times 4\times 4 =
64$  codons.

 Although there are
about 20 known codes, the most important are two of them: the
standard  code and the vertebrate mitochondrial code.

In the sequel we shall mainly have in mind the vertebrate
mitochondrial genetic code, because it is a simple one and the
others may be viewed as its slightly modified versions.  In the
vertebrate mitochondrial code, $60$ of codons are distributed on
the $20$ different amino acids and $4$ codons make termination
signal. According to experimental observations, two amino acids
are coded by six codons, six amino acids by four codons, and
twelve amino acids by two codons. This property that some amino
acids are coded by more than one codon is known as {\it genetic
code degeneracy}. This degeneracy is a very important property of
the genetic code and gives an efficient way to minimize errors
caused by mutations.

Since there is in principle a huge number (between $10^{71}$ and
$10^{84}$  (Hornos J. and Hornos Y., 1993)) of all possible
assignments between codons and amino acids,  and only a very small
number  of them is represented in living cells, it has been
permanent theoretical challenge to find an appropriate model
explaining contemporary genetic codes. There are many papers in
this direction  scattered in various journals, with theoretical
approaches based more or less on chemical, biological and
mathematical aspects of the genetic code. The first genetic model
was proposed in 1954 by physicist George Gamow (1904--1968), which
he called the diamond code. In his model codons are composed of
three nucleotides and proteins are directly synthesized at DNA:
each cavity at DNA attracts one of 20 amino acids. This is  an
overlapping code and was ruled out by analysis of correlations
between amino acids in proteins, but concept of trinucleotide
codons was correct. The next model of the genetic code was
proposed in 1957 by Crick, and is known as the comma-free code.
This model was so elegant that it was almost universally accepted.
However, an experiment in 1961 demonstrated that UUU codon codes
amino acid phenylalanine, while by this code it codes nothing.
Gamow's and Crick's models are very pretty but wrong -- living
world prefers actual codes, which are more stable with respect to
possible errors (for a popular review of the early models, see
(Hayes, 1998)).

Let us mention some models of the genetic code after deciphering
standard code. In 1966 physicist Yuri Rumer (1901--1985)
emphasized the role of the first two nucleotides in the codons
(Rumer, 1966). There are models which are based on chemical
properties of amino acids (see, e.g. (Swanson, 1984)). In some
models connections between  number of constituents of amino acids
and nucleotides and some properties of natural numbers are
investigated (see (Scherbak, 2003; Rako\v cevi\'c, 2004; Negadi,
2007) and references therein). A model based on the quantum
algebra $\mathcal{U}_q (sl(2)\oplus sl(2))$ in the $q \to 0$ limit
was proposed as a symmetry algebra for the genetic code (see
(Frappat {\it et al.}, 2001) and references therein). In a sense
this approach mimics quark model of protons and neutrons. Besides
some successes of this approach, there is a problem with rather
many parameters. There are also papers, see, e.g. (Hornos J. and
Hornos Y., 1993; Forger and Sachse, 2000; Bashford {\it et al.},
1997) starting with 64-dimensional irreducible representation of a
Lie (super)algebra and trying to connect multiplicity of codons
with irreducible representations of subalgebras arising in a chain
of symmetry breaking. Although interesting as an attempt to
describe evolution of the genetic code  these Lie algebra
approaches did not progress further.   For a very brief review of
these and some other theoretical approaches to the genetic code
one can see (Frappat {\it et al.}, 2001).




Despite of  remarkable experimental successes and some partial
theoretical descriptions, there is no simple and generally
accepted theoretical understanding of the genetic code.
 Hence, the foundation of biological coding is still an
open problem. In particular, it is not clear why genetic code
exists just in a few known ways and not in many other possible
ones. What is origin and evolution of the genetic code? Is there a
mathematical  principle behind genetic coding? We keep in mind
these and similar questions trying to find simple and general
approach, which seems to be $p$-adic ultrametricity.

\section{Ultrametric $5$-Adic  Space}

Before than consider $p$-adic properties of the genetic code in a
self-contained way we shall  recall some mathematical
preliminaries.

As a new tool to study the Diophantine equations, $p$-adic numbers
are introduced by German mathematician Kurt Hensel in 1897. They
are involved in many branches of modern mathematics. An elementary
introduction to $p$-adic numbers can be found in the book (Goueva,
1993). However, for our purposes we will use here only a small
portion of $p$-adics, mainly some finite sets of integers and
ultrametric distances between them.

Consider the  set of natural numbers

\beq \mathcal{C}_5\, [64] = \{ n_0 + n_1\, 5 + n_2\, 5^2 \,:\,\,
n_i = 1, 2, 3, 4 \}\,,   \label{2.1}\eeq where $n_i$ are digits
different from zero. This is a finite expansion to the base $ 5$.
It is obvious that $5$ is a prime number and that the set
$\mathcal{C}_5 [64]$ contains $64$ natural numbers. In the sequel
we shall often denote elements of $\mathcal{C}_5 [64]$ by their
digits to the base $5$ in the following way: $ n_0 + n_1\, 5 +
n_2\, 5^2 \, \equiv n_0\, n_1\, n_2$. Note that here ordering of
digits is the same as in the expansion, i.e this ordering is
opposite to the usual one.

It is  often important to know a distance between numbers.
Distance can be defined by a norm. On the set $\mathbb{Z}$ of
integers  there are two kinds of nontrivial norm: usual absolute
value $|\cdot|_\infty$ and $p$-adic absolute value $|\cdot|_p$ ,
where indices $\infty$ and $p$ denote real and $p$-adic case,
respectively ($p$ is any prime number). The usual absolute value
is well known from elementary mathematics and the corresponding
ordinary distance between two numbers $x$ and $y$ is $d_\infty (x,
y) = |x-y|_\infty$.

The $p$-adic absolute value is related to the divisibility of
integers by prime numbers. Difference of two integers is again an
integer. $p$-Adic distance between two integers can be understood
as a measure of  divisibility  of their difference by $p$ (the
more divisible, the shorter). By definition, $p$-adic norm of an
integer  $m \in \mathbb{Z}$, is $|m|_p = p^{-k}$, where $ k \in
\mathbb{N} \bigcup \{ 0\}$ is degree of divisibility of $m$ by
prime $p$ (i.e. $m = p^k\, m'\,, \,\, p\nmid m'$) and $|0|_p =0.$
This norm is a mapping from $\mathbb{Z}$ into non-negative
rational numbers and has the following properties:

(i) $|x|_p \geq 0, \,\,\, |x|_p =0$ if and only if $x = 0$,

(ii) $|x\, y|_p = |x|_p \,  |y|_p \,,$

(iii) $|x + y|_p \leq \, \mbox{max}\, \{ |x|_p\,, |y|_p \} \leq
|x|_p + |y|_p $ for all $x \,, y \in \mathbb{Z}$.

\noindent Because of the strong triangle inequality $|x + y|_p
\leq \, \mbox{max} \{ |x|_p\,, |y|_p \}$, $p$-adic absolute value
belongs to non-Archimedean (ultrametric) norm. One can easily
conclude that $0 \leq |m|_p \leq 1$ for any $m\in \mathbb{Z}$ and
any prime $p$.

$p$-Adic distance between two integers $x$ and $y$ is
\begin{equation}
d_p (x\,, y) = |x - y|_p \,.    \label{2.2}
\end{equation}
Since $p$-adic absolute value is ultrametric, the $p$-adic
distance (\ref{2.2}) is also ultrametric, i.e. it satisfies
inequality
\begin{equation}
d_p (x\,, y) \leq\, \mbox{max}\, \{ d_p (x\,, z) \,, d_p (z\,, y)
\} \leq d_p (x\,, z) + d_p (z\,, y) \,, \label{2.3}
\end{equation}
where $x, \, y$ and $z$ are any three integers.

 $5$-Adic distance between two numbers $a, b \in \mathcal{C}_5
\, [64]$ is

\beq d_5 (a,\, b) = |a_0 + a_1 \, 5 + a_2 \, 5^2 - b_0 - b_1 \, 5
- b_2 \, 5^2 |_5 \,,   \label{2.4} \eeq where $a_i ,\, b_i \in \{
1, 2, 3, 4\}$. When $a \neq b$ then $d_5 (a,\, b)$ may have three
different values:
\begin{itemize}
\item $d_5 (a,\, b) = 1$ if $a_0 \neq b_0$, \item $d_5 (a,\, b) =
1/5$ if $a_0 = b_0 $ and $a_1 \neq b_1$, \item $d_5 (a,\, b) =
1/5^2$ if $a_0 = b_0 \,, \,\,a_1 = b_1$ and $a_2 \neq b_2 $.
\end{itemize}
 We
see that the largest $5$-adic distance between numbers is $1$ and
it is maximum $p$-adic distance on $\mathbb{Z}$. The smallest
$5$-adic distance on the  space $\mathcal{C}_5 \, [64]$ is
$5^{-2}$. Let us also note that $5$-adic distance depends only on
the first two digits of different numbers  $a, b \in \mathcal{C}_5
\, [64]$.

If we use real (standard) distance $d_\infty (a,\, b) = |a_0 + a_1
\, 5 + a_2 \, 5^2 - b_0 - b_1 \, 5 - b_2 \, 5^2 |_\infty $, then
third digits $a_2$ and $b_2$ would play more important role than
those at the second position (i.e $a_1 \, \mbox{and} \, b_1$), and
digits $a_0$ and $b_0$ are of the smallest importance. At real
$\mathcal{C}_5 [64]$ space distances are also discrete, but take
values $1,\, 2,\, \cdots , 93$. The smallest real and the largest
$5$-adic distance are equal $1$. While real distance describes
metric of the ordinary physical space, this $p$-adic one may serve
to describe ultrametricity  of the information space.

Ultrametric space $\mathcal{C}_5 [64]$ can be viewed as 16
quadruplets with respect to the smallest $5$-adic distance, i.e.
quadruplets contain 4 elements and $5$-adic distance between any
two elements within quadruplet is $\frac{1}{25}$. In other words,
within each quadruplet elements have the first two digits equal
and third digits are different.

  With respect to  $2$-adic distance, the above quadruplets may be viewed
 as composed of two doublets: $a = a_0\, a_1\, 1$ and $b = a_0\, a_1\, 3$
 make the first doublet, and
 $c = a_0\, a_1\, 2$ and $d = a_0\, a_1\, 4$ form the second one. $2$-Adic
 distance between codons within each of these doublets is
 $\frac{1}{2}$, i.e.
 \begin{equation}
d_2 (a,\, b) = |(3 -1)\, 5^2|_2 =\frac{1}{2} \,, \, \quad \, d_2
(c,\, d) = |(4 -2)\, 5^2|_2 =\frac{1}{2} \,,   \label{2.12}
 \end{equation}
because $3-1 = 4 - 2 = 2$. By this way ultrametric space
$\mathcal{C}_5 [64]$ of 64 elements is arranged into 32 doublets.


\section{$5$-Adic Codon Space}

It is not difficult to note that ultrametric space of numbers in
$\mathcal{C}_5 [64]$ and distribution of codons in Table 2 of the
vertebrate mitochondrial code have some similarity. Already at the
first glance, one can see that both have 64 elements and that
there are quadruplets with equal the first two blocks of triples
of letters and triples of digits.

Identifying  appropriately nucleotides by digits, we obtain the
corresponding ultrametric structure of the codon space in the
vertebrate mitochondrial genetic code. We take the following
assignments between nucleotides and digits in $\mathcal{C}_5
[64]$: C (cytosine) = 1,\, A (adenine) = 2,\, T (thymine) = U
(uracil) = 3,\, G (guanine) = 4. Ordering $5$-adic numbers in
increasing way one obtains rearranged codon space and it is
presented in Table 3. There is now evident one-to-one
correspondence between codons in three-letter  notation and
three-digit $n_0\, n_1\, n_2$ number representation of ultrametric
space $\mathcal{C}_5 [64]$.

The above introduced set $\mathcal{C}_5\, [64]$ endowed by
$p$-adic distance we shall call {\it $p$-adic codon space}, i.e.
elements of $\mathcal{C}_5\, [64]$ are also codons denoted by $n_0
n_1 n_2$.

Let us now explore distances between codons and their role in
formation of the genetic code degeneration.

To this end let us again  turn to  Table 3 as a representation of
the $\mathcal{C}_5\, [64]$ codon space. Namely, we observe that
there are 16 quadruplets such that each of them has the same first
two digits. Hence $5$-adic distance between any two different
codons within a quadruplet is
\bea d_5 (a,\, b) = |a_0 + a_1 \, 5
+ a_2 \, 5^2 - a_0 - a_1 \, 5 - b_2 \, 5^2 |_5 \nn
\\= |(a_2 - b_2) \, 5^2|_5 = |(a_2 - b_2)|_5 \,\, | 5^2 |_5 =
5^{-2}\,, \label{2.11} \eea because $a_0 = b_0$, $a_1 = b_1$ and
$|a_2 - b_2|_5 = 1$. According to (\ref{2.11}) codons within every
quadruplet are at the smallest distance, i.e. they are closest
comparing to all other codons.

Since codons are composed of three ordered  nucleotides, each of
which is either a purine or a pyrimidine, it is natural to try to
quantify similarity inside purines and pyrimidines, as well as
distinction between elements from these two groups of nucleotides.
Fortunately there is a tool, which is again related to the
$p$-adics, and now it is $2$-adic distance. One can easily see
that  $2$-adic distance between pyrimidines  C and U is $d_2 (1,
3) = |3 - 1|_2 = 1/2$ as the distance between purines  A and G,
namely $d_2 (2, 4) = |4 - 2|_2 = 1/2$. However $2$-adic distance
between C and A or G as well as distance between U and A or G is
$1$ (i.e. maximum).

One can now look at Table 3 as a system of 32 doublets. Thus 64
codons are clustered by a very regular way into 32 doublets. Each
of 21 subjects (20 amino acids and 1 termination signal) is coded
by one, two or three doublets. In fact, there are two, six and
twelve amino acids coded by three, two and one doublet,
respectively. Residual two doublets code termination signal.

Note that 2  doublets code 2 amino acids (Ser and Leu) which are
already coded by 2 quadruplets, thus amino acids Serine and
Leucine are coded by 6 codons (3 doublets).

To have a more complete picture of the genetic code it is useful
to consider possible distances between codons of different
quadruplets as well as between different doublets. Also, we
introduce distance between quadruplets or between doublets,
especially when distances between their codons have the same
value. Thus $5$-adic distance between any two quadruplets  in the
same column is $1/5$, while such distance between  other
quadruplets is $1$. $5$-Adic distance between doublets coincides
with distance between quadruplets, and this distance is
$\frac{1}{5^2}$ when doublets are within the same quadruplet.

The $2$-adic distances between codons, doublets and  quadruplets
are more complex. There are three basic cases: \begin{itemize}
\item codons differ only in one digit, \item codons differ in two
digits, \item codons differ in all three digits.
\end{itemize}
In the first case, $2$-adic distance can be $\frac{1}{2}$  or $1$
depending whether difference between digits is $2$ or not,
respectively.

Let us now look at $2$-adic distances between doublets coding
Leucine and also between doublets coding Serine. These are two
cases of amino acids coded by three doublets. One has the
following distances:
\begin{itemize}
\item $d_2 (332, 132) = d_2 (334, 134) = \frac{1}{2}$  for
Leucine, \item $d_2 (311, 241) = d_2 (313, 243) = \frac{1}{2}$ for
Serine.
\end{itemize}

If we use usual distance  between codons, instead of $p$-adic one,
then we would observe that two synonymous codons are very far (at
least 25 units), and that those which are close code different
amino acids. Thus we conclude that not usual metric but
ultrametric is inherent to codon space.

How degeneracy of the genetic code is connected with $p$-adic
distances between codons? The answer is in the following basic
{\bf $p$-adic degeneracy principle}: {\it Amino acids are coded by
doublets of codons, where a doublet contains two nucleotides of
the smallest $(\frac{1}{5})$ $5$-adic distance and $\frac{1}{2}\,$
$2$-adic distance}. Here $p$-adic distance plays a role of
similarity: the closer, the more similar. Taking into account the
standard genetic code, there is a slight violation of this
principle. Now it is worth noting that in modern particle physics
just broken of the fundamental gauge symmetry gives its standard
model. There is a sense to introduce a new principle (let us call
it {\bf reality principle}): {\it Reality is realization of some
broken fundamental principles}. It seems that this principle is
valid not only in physics but also in all sciences. In this
context modern genetic code, especially the standard genetic code,
is an evolutionary broken the above $p$-adic degeneracy principle.

\begin{table}
 {{\bf Table 3.} The $p$-adic vertebrate mitochondrial genetic
code. \vskip3mm } \centerline{ {\begin{tabular}{|l|l|l|l|}
 \hline \ & \ & \ & \\
  111 \, CCC \, Pro &   211 \, ACC \, Thr  &  311 \, UCC \, Ser &  411 \, GCC \, Ala  \\
  112 \, CCA \, Pro &   212 \, ACA \, Thr  &  312 \, UCA \, Ser &  412 \, GCA \, Ala  \\
  113 \, CCU \, Pro &   213 \, ACU \, Thr  &  313 \, UCU \, Ser &  413 \, GCU \, Ala  \\
  114 \, CCG \, Pro &   214 \, ACG \, Thr  &  314 \, UCG \, Ser &  414 \, GCG \, Ala  \\
 \hline \  & \  &  \ & \ \\
  121 \, CAC \, His &   221 \, AAC \, Asn  &  321 \, UAC \, Tyr &  421 \, GAC \, Asp  \\
  122 \, CAA \, Gln &   222 \, AAA \, Lys  &  322 \, UAA \, Ter &  422 \, GAA \, Glu  \\
  123 \, CAU \, His &   223 \, AAU \, Asn  &  323 \, UAU \, Tyr &  423 \, GAU \, Asp  \\
  124 \, CAG \, Gln &   224 \, AAG \, Lys  &  324 \, UAG \, Ter &  424 \, GAG \, Glu  \\
 \hline \  & \  & \  &   \\
  131 \, CUC \, Leu &   231 \, AUC \, Ile  &  331 \, UUC \, Phe &  431 \, GUC \, Val  \\
  132 \, CUA \, Leu &   232 \, AUA \, Met  &  332 \, UUA \, Leu &  432 \, GUA \, Val  \\
  133 \, CUU \, Leu &   233 \, AUU \, Ile  &  333 \, UUU \, Phe &  433 \, GUU \, Val  \\
  134 \, CUG \, Leu &   234 \, AUG \, Met  &  334 \, UUG \, Leu &  434 \, GUG \, Val  \\
 \hline \ & \   & \  &   \\
  141 \, CGC \, Arg &   241 \, AGC \, Ser  &  341 \, UGC \, Cys &  441 \, GGC \, Gly  \\
  142 \, CGA \, Arg &   242 \, AGA \, Ter  &  342 \, UGA \, Trp &  442 \, GGA \, Gly  \\
  143 \, CGU \, Arg &   243 \, AGU \, Ser  &  343 \, UGU \, Cys &  443 \, GGU \, Gly  \\
  144 \, CGG \, Arg &   244 \, AGG \, Ter  &  344 \, UGG \, Trp &  444 \, GGG \, Gly  \\
\hline
\end{tabular}}{}}
\end{table}

Let us now turn to   Table 2. We observe that this table can be
regarded as a big rectangle divided into 16 equal smaller
rectangles: 8 of them are quadruplets which one-to-one correspond
to 8 amino acids, and other 8 rectangles are divided into 16
doublets coding 14 amino acids and termination (stop) signal (by
two doublets at different places). However there is no
 manifest symmetry in distribution of these quadruplets and
doublets.

In order to get a symmetry we have rewritten this standard table
into new one  by rearranging 16 rectangles. As a result we
obtained  Table 3 which exhibits a symmetry with respect to the
distribution of codon quadruplets and codon doublets. Namely, in
our table quadruplets and doublets form separately two figures,
which are symmetric with respect to the mid vertical line (a
left-right symmetry), i.e. they are invariant under interchange $C
\leftrightarrow G$ and $A \leftrightarrow U$ at the first position
in codons at all horizontal lines. The observed left-right
symmetry is now invariance under the corresponding transformations
$1 \leftrightarrow 4$ and $2 \leftrightarrow 3$. In other words,
at each horizontal line one can perform {\it doublet}
$\leftrightarrow$ {\it doublet} and {\it quadruplet}
$\leftrightarrow$ {\it quadruplet} interchange around vertical
midline. Recall that also  DNA is symmetric under simultaneous
interchange of complementary nucleotides $C \leftrightarrow G$ and
$A \leftrightarrow T$ between its strands. All doublets in this
table form a nice figure which looks like letter $\mathbb{T}$.


Note that  the above invariance leaves also unchanged polarity and
hydrophobicity  of the corresponding amino acids in all but three
cases: Asn $ \leftrightarrow $ Tyr, Arg $ \leftrightarrow $ Gly,
and Ser $ \leftrightarrow $ Cys.

It is also worth noting that four nucleotides are related to prime
number $5$ by their correspondence to the four nonzero digits $(1,
2, 3, 4)$ of $p = 5$. It is unappropriate to use the digit $0$ for
a nucleotide because it leads to non-uniqueness  in representation
of the codons by natural numbers. For example,  $123 = 123000$ as
numbers, but $123$ would represent one and $123000$ two codons.
This is also a reason why we do not use $4$-adic representation
for codons, since it would contain a nucleotide presented by digit
$0$.  One can use $0$ as a digit to denote absence of any
nucleotide.

\section{$5$-Adic Amino Acids Space}

\begin{table}
{{\bf Table 4.}  $20$ standard amino acids with assigned
corresponding numbers. \vskip3mm }
  \centerline{ {\begin{tabular}{|l|l|l|l|}
 \hline \ & \ & \ & \\
 11 \, Proline &  21 \, Threonine  & 31 \, Serine & 41 \, Alanine  \\
  \hline \  & \  &  \ & \ \\
 12 \, Histidine &  22 \, Asparagine  & 32 \, Tyrosine & 42 \, Aspartate  \\
  \hline \  & \  & \  &   \\
 13 \, Leucine &  23 \, Isoleucine  & 33 \, Phenynalanine & 43 \, Valine \\
  \hline \ & \   & \  &   \\
 14 \, Arginine &  24 \, Lysine  & 34 \, Cysteine & 44 \, Glycine  \\
 \hline \ & \   & \  &   \\
 1\, \, Glutamine & \ 2\, \, Methionine & 3\, Tryptophan & 4\, \, Glutamate \\
 \hline
\end{tabular}}{}}
\end{table}


At Table 4 we assigned numbers $x_0 x_1 \equiv x_0 + x_1 \, 5 $ to
 $16$ amino acids which are assumed to be present in dinucleotide
coding epoch, and $x_0 = 1, 2, 3, 4$ is attached to  four late
amino acids which were added during trinucleotide coding. Having
these $5$-adic numbers for amino acids one can consider  distance
between codons and amino acids: there are $23$ codon doublets
which are at $\frac{1}{25}$ $\, 5$-adic distance with the
corresponding $15$ amino acids, i.e. these codons within doublets
and related amino acids are at the same $5$-adic distance.

\section{Possible Evolution of the Genetic Code}

There are two types of evolution of the genetic code: 1) evolution
of the codon space and 2) evolution of amino acids sector with
fixed trinucleotide codon space. We shall discuss mainly the first
type of evolution.

The origin  and early evolution of the genetic code are among the
most interesting and important  investigations  related to the
origin and  evolution of the life. However, since there are no
concrete fossils  from that early period, it gives rise to many
speculations. Nevertheless, one can  hope that some of the
hypotheses may be tested looking for the corresponding traces in
the contemporary genomes.

It seems natural to consider biological evolution as an adaptive
development of simpler living systems to more complex ones.
Namely, living organisms are open systems in permanent interaction
with environment. Thus the evolution can be modelled by a system
with given initial conditions and guided by some internal rules
taking into account environmental factors.

We are going now to conjecture on the evolution of the genetic
code using our p-adic approach to the codon space, and assuming
that preceding  codes used simpler codons and  older amino acids.

\begin{table}
{{\bf Table 5.}   $5$-Adic  system including digit $0$, and
containing single nucleotide, dinucleotide and trinucleotide
codons.

 \vskip3mm \label{Tab:07}}\centerline{ {\begin{tabular}{|l|l|l|l|l|}
 \hline \  & \   & \   & \ &  \\
 000   & 100 \bf C & 200 \bf A & 300 \bf U & 400 \bf G  \\
 \hline \  & \   &  \  & \ &  \\
 010 & 110 \bf CC & 210 \bf AC & 310 \bf UC & 410 \bf GC  \\
 020 & 120 \bf CA & 220 \bf AA & 320 \bf UA & 420 \bf GA  \\
 030 & 130 \bf CU & 230 \bf AU & 330 \bf UU & 430 \bf GU  \\
 040 & 140 \bf CG & 240 \bf AG & 340 \bf UG & 440 \bf GG \\
  \hline \  & \   &  \  & \ &  \\
 001 & 101 & 201 & 301 & 401  \\
  \hline \  & \   &  \  & \ &  \\
 011 & 111 \bf CCC & 211 \bf ACC & 311 \bf UCC & 411 \bf GCC  \\
 021 & 121 \bf CAC & 221 \bf AAC & 321 \bf UAC & 421 \bf GAC  \\
 031 & 131 \bf CUC & 231 \bf AUC & 331 \bf UUC & 431 \bf GUC  \\
 041 & 141 \bf CGC & 241 \bf AGC & 341 \bf UGC & 441 \bf GGC  \\
 \hline \  & \  &  \   & \ &  \\
 002 & 102 & 202 & 302 & 402  \\
  \hline \  & \   &  \  & \ &  \\
 012 & 112 \bf CCA & 212 \bf ACA & 312 \bf UCA & 412 \bf GCA  \\
 022 & 122 \bf CAA & 222 \bf AAA & 322 \bf UAA & 422 \bf GAA  \\
 032 & 132 \bf CUA & 232 \bf AUA & 332 \bf UUA & 432 \bf GUA  \\
 042 & 142 \bf CGA & 242 \bf AGA & 342 \bf UGA & 442 \bf GGA \\
 \hline \  & \   & \   & \ &  \\
 003 & 103 & 203 & 303 & 403  \\
  \hline \  & \   &  \  & \ &  \\
 013 & 113 \bf CCU & 213 \bf ACU & 313 \bf UCU & 413 \bf GCU  \\
 023 & 123 \bf CAU & 223 \bf AAU & 323 \bf UAU & 423 \bf GAU  \\
 033 & 133 \bf CUU & 233 \bf AUU & 333 \bf UUU & 433 \bf GUU  \\
 043 & 143 \bf CGU & 243 \bf AGU & 343 \bf UGU & 443 \bf GGU  \\
 \hline \  & \   & \   & \ &  \\
 004 & 104 & 204 & 304 & 404  \\
  \hline \  & \   &  \  & \ &  \\
 014 & 114 \bf CCG & 214 \bf ACG & 314 \bf UCG & 414 \bf GCG  \\
 024 & 124 \bf CAG & 224 \bf AAG & 324 \bf UAG & 424 \bf GAG  \\
 034 & 134 \bf CUG & 234 \bf AUG & 334 \bf UUG & 434 \bf GUG  \\
 044 & 144 \bf CGG & 244 \bf AGG & 344 \bf UGG & 444 \bf GGG  \\
\hline
\end{tabular}}{}}
\vskip0.3cm Ignoring numbers which contain digit 0 in front of any
1, 2, 3 or 4,  one has one-to-one correspondence between 1-digit,
2-digits, 3-digits numbers and single nucleotides, dinucleotides,
trinucleotides, respectively. It seems that evolution of codons
has followed transitions: single nucleotides $\to$ dinucleotides
$\to$ trinucleotides.

\end{table}

\begin{table}
{{\bf Table 6.} {Temporal appearance of the 20 standard amino
acids} (Trifonov, 2004).

\vskip3mm \label{Tab:03}} {\centerline{\begin{tabular}{|l|l|l|l|}
\hline \ & \ & \ & \\
(1) Gly & (2) Ala & (3) Asp & (4) Val\\
\hline \ & \ & \ & \\
(5) Pro & (6) Ser & (7) Glu & (8) Leu\\
\hline \ & \ & \ & \\
(9) Thr & (10) Arg & (11) Ile & (12) Gln \\
\hline \ & \ & \ & \\
(13) Asn & (14) His & (15) Lys & (16) Cys\\
\hline \ & \ & \ & \\
(17) Phe & (18) Tyr & (19) Met & (20) Trp\\
\hline
\end{tabular}}{}}
\end{table}


Consider general $p$-adic codon  space $\mathcal{C}_5 \, \big[
(p-1)^m \big]$ which has two parameters: $p$ - related to $p-1$
building blocks, and $m$ - multiplicity of the building blocks in
codons. Then
\begin{itemize}
\item Case $\mathcal{C}_2 \, \big[ 1  \big]$ is a trivial one and
useless for a primitive code.

\item Case $\mathcal{C}_3 \, \big[  2^m \big]$ with $m =1, 2, 3$
does not seem to be realistic.

\item Case $\mathcal{C}_5 \, \big[ 4^m  \big]$ with $m = 1, 2, 3$
offers a possible pattern to consider evolution of the genetic
code. Namely, the codon space could evolve in the following way:
$\mathcal{C}_5 \, \big[ 4  \big] \to \mathcal{C}_5 \, \big[ 4^2
\big] \to \mathcal{C}_5 \, \big[ 4^3  \big] = \mathcal{C}_5\,
[64].$
\end{itemize}

According to  Table 5 the primary code, containing codons in the
single nucleotide form (C, A, U, G), encoded the first four amino
acids (see Table 6): Gly, Ala, Asp and Val. From the last column
of Table 4 we conclude that the connection between digits and
amino acids is: $1 \to Ala,\, 2 \to Asp,\, 3 \to Val,\, 4 \to
Gly$. In the primary code these digits occupied the first position
in the $5$-adic expansion (Table 5), and at the next step, i.e.
$\mathcal{C}_5 \, \big[ 4 \big] \to \mathcal{C}_5 \, \big[ 4^2
\big]$, they moved to the second position adding digits $1, 2, 3,
4$ in front of  each of them.

\begin{table}
{\bf Table  7.} {The dinucleotide genetic code based on the
$p$-adic codon space $\mathcal{C}_5 \, [ 4^2 ]$. }  \vspace{0.4cm} \\
\centerline{ {\begin{tabular}{|l|l|l|l|}
 \hline \ & \ & \ & \\
 11 \, CC \, Pro &  21 \, AC \, Thr  & 31 \, UC \, Ser & 41 \, GC \, Ala  \\
  \hline \  & \  &  \ & \ \\
 12 \, CA \, His &  22 \, AA \, Asn  & 32 \, UA \, Tyr & 42 \, GA \, Asp  \\
  \hline \  & \  & \  &   \\
 13 \, CU \, Leu &  23 \, AU \, Ile  & 33 \, UU \, Phe & 43 \, GU \, Val \\
  \hline \ & \   & \  &   \\
 14 \, CG \, Arg &  24 \, AG \, Ser  & 34 \, UG \, Cys & 44 \, GG \, Gly  \\
 \hline
\end{tabular}}{}}
\vskip0.3cm Note that this code encodes 15 amino acids without
stop codon, but encodes Serine twice.
\end{table}

In $\mathcal{C}_5 \, \big[ 4^2  \big]$ one has 16 dinucleotide
codons which can code up to 16 new amino acids. Addition of the
digit $4$ in front of already existing codons $1, 2, 3, 4$ leaves
their meaning unchanged, i.e. $41 \to Ala, \, 42 \to Asp,\, 43 \to
Val, \, 44 \to Gly.$ Adding digits $3, 2, 1$ in front of the
primary $1, 2, 3, 4$ codons one obtains 12  possibilities for
coding some new amino acids. To decide which amino acid was
encoded by which of 12 dinucleotide codons, we use as a criterion
their immutability  in the trinucleotide coding on the
$\mathcal{C}_5 \, \big[ 4^3 \big]$ space. This criterion  assumes
that amino acids encoded earlier are more fixed than those encoded
later. According to this criterion we decide in favor of the first
row in each rectangle of Table 3 and result is presented in  Table
7.

Transition from  dinucleotide to trinucleotide codons occurred by
attaching nucleotides $1, 2, 3, 4$ at the third position, i. e.
behind each dinucleotide. By this way one obtains new codon space
$\mathcal{C}_5 \, \big[ 4^3  \big] = \mathcal{C}_5\, [64]$, which
is significantly enlarged and  provides a pattern to generate
known genetic codes. This codon space $ \mathcal{C}_5\, [64]$
gives possibility to realize at least three general properties of
the modern code: 

 (i) encoding of more than 16 amino acids,

(ii) diversity of codes,

(iii) stability of the gene expression.

\bigskip
Let us give some relevant clarifications.

(i) For functioning of contemporary living organisms it is
necessary to code at least 20 standard (Table 1)  and 2
non-standard amino acids (selenocysteine and pyrrolysine).
Probably these 22 amino acids are also sufficient building units
for biosynthesis of all necessary contemporary proteins. While $
\mathcal{C}_5 \, \big[ 4^2 \big]$ is insufficient, the codon space
$ \mathcal{C}_5 \, \big[ 4^3 \big]$ offers approximately three
codons per one amino acid.

(ii) The  standard  code was deciphered around 1966 and was
thought to be universal, i. e., common to all organisms. When the
human mitochondrial code was discovered in 1979, it gave rise to
believe that the code is not frozen and that there are also some
other codes which are mutually different. According to later
evidences, one can say that there are about 20 slightly different
mitochondrial and nuclear codes (for a review, see (Knight {\it et
al.}, 2001; Osawa {\it et al.}, 1992) and references therein).
Different codes have some codons with different meaning. So, in
the standard genetic code there are the following changes in Table
3:
\begin{itemize}\item 232 (AUA):
Met $\rightarrow$ Ile, \item 242 (AGA) and 244 (AGG): Ter
$\rightarrow$ Arg, \item 342 (UGA): Trp $\rightarrow$ Ter.
\end{itemize}

(iii) Each of the 20 codes is degenerate and degeneration provides
their stability against possible mutations. In other words,
degeneration helps to minimize codon errors.

Genetic codes based on single nucleotide and dinucleotide codons
were mainly directed to code amino acids with rather different
properties. This may be the reason why amino acids Glu and Gln are
not coded in dinucleotide code (Table 7), because they are similar
to Asp and Asn, respectively.   However, to become almost optimal,
trinucleotide codes have taken into account structural and
functional similarities of amino acids.

We presented here a hypothesis on the genetic code evolution
taking into account possible codon evolution, from 1-nucleotide to
3-nucleotide, and amino acids temporal appearance. This scenario
may be extended to the cell evolution, which probably should be
considered as a coevolution of all its main ingredients (for an
early idea of the coevolution, see (Wong, 1975)).

\section{Concluding Remarks}


There are two aspects of the genetic code related to:

(i)  multiplicity of codons which code the same amino acid,

(ii) concrete assignment of codon multiplets to particular amino
acids.

The above presented $p$-adic approach gives quite satisfactory
description of the aspect (i). Ultrametric behavior of $p$-adic
distances between elements of  the $\mathcal{C}_5 \,[64]$ codon
space radically differs from the usual ones. Quadruplets and
doublets of codons  have natural explanation within $5$-adic and
$2$-adic nearness. Degeneracy of the genetic code in the form of
doublets, quadruplets and sextuplets is direct consequence of
$p$-adic ultrametricity between codons. $p$-Adic $\mathcal{C}_5\,
[64]$ codon space is our theoretical pattern to consider all
variants of the genetic code: some codes are direct representation
of $\mathcal{C}_5 \, [64]$ and the others are its slight
evolutional modifications.

(ii) Which amino acid corresponds to which doublet of codons? An
answer to this question should be expected from connections
between physicochemical properties of  amino acids and anticodons.
Namely, enzyme aminoacyl-tRNA synthetase links specific tRNA
anticodon and related amino acid. Thus there is no direct
interaction between amino acids and trinucleotide codons, as it
was believed for some time in the past. However, from our $p$-adic
analysis follows that at an epoch of dinucleotide codons
connection of codons and amino acids should be direct. Namely, at
that time $p$-adic distance between dinucleotide codons and some
amino acids was zero.



%
%

Note that there are in general $4!$ ways to assign digits $1, 2,
3, 4$ to nucleotides C, A, U, G. After an analysis of all 24
possibilities, we have taken C = 1, \, A = 2,\, U = T = 3,\, G = 4
as a quite appropriate choice. In addition to various properties
already presented in this paper it exhibits also complementarity
of nucleotides in the DNA double helix  by relation C + G = A + T
= 5.

One can  express many above considerations of $p$-adic information
theory in linguistic terms and investigate possible linguistic
aspects.

In this paper we have employed $p$-adic distances to measure
similarity between codons, which have been used to describe
degeneracy of the genetic code and to propose its evolution. It is
worth noting that in other contexts $p$-adic distances can be
interpreted in quite different meanings. For example, $3$-adic
distance between cytosine and guanine is $d_3 (1, 4) =
\frac{1}{3}$, and between adenine and thymine $d_3 (2, 3) =1$.
This $3$-adic distance seems to be natural to relate to hydrogen
bonds between complements in DNA double helix: the smaller
distance, the stronger hydrogen bond. Recall that  C-G and A-T are
bonded by 3 and 2 hydrogen bonds, respectively.

The translation of codon sequences into proteins is highly an
information-processing phenomenon. $p$-Adic information modelling
presented in this paper offers a new approach to systematic
investigation of ultrametric aspects of  DNA and RNA sequences,
the genetic code and the world of proteins. It can be embedded in
computer programs to explore $p$-adic side of the genome and
related subjects.

The above considerations and obtained results may be viewed as
contribution to foundation of  $p$-adic theory of the genetic
code, but also to theory of $p$-adic information.

\section*{Acknowledgements}

The work  on this paper was partially supported by the Ministry of
Education and Science, Serbia, contracts 173052 and 174012. The
author would like to thank Alexandra Yu. Dragovich and Miloje
Rako\v cevi\'c for fruitful discussions of various aspects of the
genetic code. The author also  appreciate activity of Sultan
Tarlaci, Chief-Editor, and  Tidjani Negadi, editor of this issue,
 on edition of the special issue of the journal NeuroQuantology
 devoted to the modern developments in understanding the genetic code and
 related topics.


\section*{References}


\vskip.3cm \noindent  Avetisov VA,   Bikulov AKh, Kozyrev SV and
 Osipov VA.  $p$-Adic Models of Ultrametric Diffusion Constrained by
 Hierarchical Energy Landscape.  J Phys A: Math  Gen 2002; 35 (2):
177--189. arXiv:cond-mat/0106506.

\vskip.3cm \noindent
 Bashford JD, Tsohantjis I and  Jarvis PD.
 Codon and Nucleotide Assignments in a Supersymmetric Model of
the Genetic Code. Phys Lett A 1997; 233: 481--488.

\vskip.3cm \noindent
  Brekke L and  Freund PGO.  $p$-Adic Numbers in Physics.
Phys Reports 1993; 233:  1--66.

\vskip.3cm \noindent
 Crick F.  The Origin of the Genetic Code. J Mol Biol 1968; 38:
367--379.

\vskip.3cm \noindent
   Dragovich B.   $p$-Adic and Adelic Quantum Mechanics. Proc VA
Steklov Inst Math 2004; 245: 72--85. arXiv:hep-th/0312046.

\vskip.3cm \noindent
  Dragovich B.  $p$-Adic and Adelic Cosmology: $p$-Adic
Origin of Dark Energy and Dark Matter. $p$-Adic Mathematical
Physics. AIP Conference Proceedings 2006; 826: 25--42.
arXiv:hep-th/0602044.

\vskip.3cm \noindent
 Dragovich B. Genetic Code and Number Theory. To apper in book
"Modern Topics in Science". arXiv:0911.4014v1 [q-bio.OT]

\vskip.3cm \noindent
 Dragovich B and  Dragovich A.  $p$-Adic Degeneracy of the
Genetic Code. SFIN A 2007; 20 (1): 179--188.
arXiv:0707.0764v1[q-bio.GN].

\vskip.3cm \noindent
  Dragovich B and  Dragovich A.   A $p$-Adic Model of DNA
Sequence and Genetic Code. $p$-Adic Numbers, Ultrametric Analysis
and Applications 2009; 1 (1): 34--41. arXiv:q-bio.GN/0607018v1.

\vskip.3cm \noindent
 Dragovich B and  Dragovich A.  $p$-Adic Modelling of the Genome
and the Genetic Code. Computer Journal 2010; 53 (4): 432--442.
arXiv:0707.3043v1 [q-bio.OT].

\vskip.3cm \noindent
 Dragovich B, Khrennikov AYu, Kozyrev SV and Volovich IV. On
$p$-Adic Mathematical Physics. $p$-Adic Numbers, Ultrametric
Analysis and Applications 2009; 1 (1): 1--17.

\noindent arXiv:0904.4205v1[math-ph].

\vskip.3cm \noindent
 Finkelshtein AV and Ptitsyn OB. Physics of Proteins. Academic
Press, London, 2002.

\vskip.3cm \noindent
 Forger M and  Sachse S.  Lie Superalgebras
 and the Multiplet Structure of the Genetic Code I: Codon
 Representations. J Math Phys 2000; 41 (8): 5407--5422.  arXiv:math-ph/9808001.

\vskip.3cm \noindent
 Frappat L,  Sciarrino A and  Sorba P.  Crystalizing the
Genetic Code. J Biol Phys 2001; 27: 1--38. arXiv:physics/0003037.

\vskip.3cm \noindent
 Gouvea FQ.  $p$-Adic Numbers: An Introduction. (Universitext),
Springer, Berlin, 1993.

\vskip.3cm \noindent
 Hayes B.  The Invention of the Genetic Code. American Scientist
1998; 86 (1): 8--14.

\vskip.3cm \noindent
 Hornos JEM and  Hornos YMM.  Algebraic Model for the Evolution
of the Genetic Code. Phys Rev Lett 1993;
 71: 4401--4404.


\vskip.3cm \noindent
 Khrennikov A. Information Dynamics in Cognitive, Psychological,
Social and Anomalous Phenomena. Kluwer AP, Dordrecht, 2004.

\vskip.3cm \noindent
 Khrennikov A  and Kozyrev S.  Genetic Code on a Diadic Plane.
Physica A: Stat Mech Appl 2007; 381: 265--272.
arXiv:q-bio/0701007.

\vskip.3cm \noindent
 Knight RD, Freeland SJ and Landweber LF. Rewiring the Keyboard:
Evolvability of the Genetic Code. Nat Rev Genet 2001; 2: 49--58.

\vskip.3cm \noindent
 Negadi T. The Genetic Code Multiplet Structure, in One Number.
Symmetry: Culture and Science 2007; 18 (2-3): 149--160.
arXiv:0707.2011v1 [q-bio.OT].

\vskip.3cm \noindent
 Osawa S, Jukes TH, Watanabe K and Muto A.
 Recent Evidence for Evolution of the Genetic Code. Microb Rev
1992; \textbf{56} (1): 229--264.

\vskip.3cm \noindent
  Rako\v cevi\'c MM.  A Harmonic Structure of the Genetic Code. J
Theor Biol 2004; 229:  221--234.

\vskip.3cm \noindent
  Rammal R,  Toulouse G and  Virasoro MA. Ultrametricity for
Physicists. Rev  Mod Phys 1986; 58:  765--788.

\vskip.3cm \noindent
 Rumer YuB.  On Systematization of Codons in the Genetic Code.
Doklady Acad Nauk USSR 1966; 167 (6): 1393--1394.

\vskip.3cm \noindent
  Shcherbak VI.  Arithmetic Inside the Universal Genetic Code.
Biosystems 2003; 70:  187--209.

\vskip.3cm \noindent
 Swanson R.  A Unifying Concept for the Amino Acid Code. Bull
Math Biol 1984; 46 (2): 187--203.

\vskip.3cm \noindent
 Trifonov EN.  The Triplet Code From First Principles. J Biomol
Struc  Dynam  2004;  22 (1): 1--11.

\vskip.3cm \noindent
  Vladimirov  VS,  Volovich IV and Zelenov EI.  $p$-Adic Analysis
and Mathematical Physics. World Scientific, Singapore, 1994.

\vskip.3cm \noindent
  Watson JD,   Baker TA,  Bell SP,  Gann A, Levine M  and Losick
R.   Molecular Biology of the Gene. CSHL Press, Benjamin Cummings,
San Francisco, 2004.

\vskip.3cm \noindent
 Wong JTF.  A Co-Evolution Theory of the Genetic Code. Proc Nat
Acad Sci USA 1975; 72: 1909--1912.




\end{document}